\documentclass[singlecolumn,superscriptaddress]{revtex4}
\usepackage{empheq}
\usepackage{amsmath}
\usepackage{amssymb}
\usepackage{dsfont}
\usepackage{graphicx}
\usepackage{hyperref}
\usepackage{stackrel}
\usepackage{color}
\usepackage{comment}

\def \ee{\end{equation}}
\def \be{\begin{equation}}
\def \eea{\end{eqnarray}}
\def \bea{\begin{eqnarray}}

\begin{document}

\title{A hidden constraint on the Hamiltonian formulation of relativistic worldlines}

\author{Benjamin Koch}
%\email{maxbanados@fis.puc.cl}
\affiliation{Pontificia Universidad Cat\'olica de Chile \\ Instituto de F\'isica, Pontificia Universidad Cat\'olica de Chile, \\
Casilla 306, Santiago, Chile}
\author{Enrique Mu\~noz}
%\email{gduring@fis.puc.cl}
\affiliation{Pontificia Universidad Cat\'olica de Chile \\ Instituto de F\'isica, Pontificia Universidad Cat\'olica de Chile, \\
Casilla 306, Santiago, Chile}
%\author{Ignacio A. Reyes}
%\email{iareyes@uc.cl}
%\affiliation{Pontificia Universidad Cat\'olica de Chile \\ Instituto de F\'isica, Pontificia Universidad Cat\'olica de Chile, \\
%Casilla 306, Santiago, Chile}
%\affiliation{Insitut f{\"u}r Theoretische Physik und Astrophysik, Julius-Maximilians-Universit{\"a}t W{\"u}rzburg, Am Hubland, 97074, Germany}

\begin{abstract}
Gauge theories with general covariance are particularly reluctant to quantization.
We discuss the example of the Hamiltonian formulation of the relativistic point particle that, despite its apparent simplicity,
is of crucial importance since a number of point particle systems can be cast into this form on a higher dimensional Rindler background, as recently pointed out by Hojman.
It is shown that this system can be equipped with a hidden local, symmetry generating, constraint which 
on the one hand does not bother the classical evolution and on the other hand 
simplifies the realization of the path integral quantization. Even though the positive impact
of the hidden symmetry is more evident in the Lagrangian version of the theory, it 
is still present through the suggested Hamiltonian constraint.
\end{abstract}

\maketitle

\tableofcontents
%%%%%%%%%%%%%%%%
\section{Introduction}

Symmetries have been the guiding principle of theoretical physics throughout centuries.
In particular local symmetries have shown to be particularly useful for
the description of fundamental interactions. The description of all
four known forces of nature: electromagnetism,
weak interaction, strong interaction, and gravitation have been cast in this language.
However, the last member of this illustrious list poses serious problems,
when it comes to a quantum formulation of gravity.
Numerous attempts have been made to solve the problem but, up to now,
no conclusion could be reached (see \cite{Kiefer:2005uk} for a review). 
Clearly, one needs to better understand the quantization of this theory, based
on a particularly beautiful and complicated symmetry called general covariance.
Since the rich structure of the full gravitational covariant system appears to be too
complicated to tackle the problem directly, it seems instead that a wiser strategy is to learn more about
general covariance in simpler systems.

Probably, the simplest system with general covariance is the relativistic point particle 
(RPP)~\cite{Teitelboim:1982,Henneaux:1982ma,Redmount:1990mi,Fradkin:1991ci,Padmanabhan:1994}.
Even though the quantization of the RPP was achieved in the Hamiltonian version
of this theory,  attempts to quantize the Lagrangian version of the action of the relativistic
point particle
\be\label{actL}
S= \int d\lambda \sqrt{\frac{d x^\mu}{d \lambda} \frac{d x_\mu}{d \lambda} }
\ee
lead to deep problems and inconsistencies. A Lagrangian  
action is most naturally quantized in a path integral (PI) approach, so in what follows
we shall refer to this quantization method.
The attitudes towards this problem that can be found in the literature are:
\begin{itemize}
\item[a)] declare the Lagrangian action (\ref{actL}) to be wrong or at least inadequate 
 for the purpose of quantization, and stick to the Hamiltonian version. This Hamiltonian version
 of the action was first formulated in~\cite{Brink:1976,Brink:1977,Fradkin:1991ci};
\item[b)] stick to the straight forward PI quantization of  (\ref{actL}) and try to 
tackle the arising problems by a re-definition of the usual interpretation of probability,
the super-probability~\cite{Jizba:2008,Jizba:2010pi};
\item[c)] study the system on special manifolds~\cite{Fukutaka:1986ps} or use approximations~\cite{Padmanabhan:1994};
\item[d)] realize that the action has a hidden symmetry, which, when factored out of the 
PI solves the inconsistencies and the quantization works just fine, giving the expected results.
The factorization was shown to work in a formal Fadeev Popov construction~\cite{Koch:2017bvv} and in a purely geometrical 
approach~\cite{Koch:2017nha}.
\end{itemize}
Clearly the option d) is the most favorable solution, due to its straightforward conceptual interpretation and its effectiveness at yielding
the correct answers.
The key point of this solution was that it treated two paths which are connected by
 the symmetry of local velocity rotations
as physically equivalent. Factoring out those paths from the naive PI calculation
solved the aforementioned pathologies. When one tries to connect this astonishing 
result with the PI quantization of the corresponding Hamiltonian system, several
questions arise:\\
 Is there a corresponding additional symmetry in the Hamiltonian system?
 If it exists, can this additional symmetry be written in terms of a (local) constraint?
 If this can be done, how does this constrain affect the PI construction of the Hamiltonian system?

On the following pages, those questions will be answered.
\\\\
The paper is organized as follows: We first start with a thorough discussion of the different criteria that a Hamiltonian formulation
of the path integral for the RPP must meet, in order to capture the symmetries involved. Then, we discuss the consequences of treating
a Hamiltonian formulation that does not take into account the hidden symmetry. 
Finally, we propose
an ansatz to involve this symmetry explicitly in the action, via a suitable constraint and a corresponding Lagrange multiplier. Moreover, by
explicit calculation of the path integral, we show that this constraint does indeed allow us to recover the correct result for the RPP propagator. At last,
we present the conclusions and possible extensions of our formulation to other physical systems of current interest.

%%%%%%%%%%%%%%%%%%%%
\section{Hamiltonian PI for the RPP}

The aim is to formulate a Hamiltionian theory for the relativistic point particle
that meets several criteria.

Let us summarize those in the following table, where the central column describes
the criteria and the right column the motivation.

\begin{table}[h!]
\begin{tabular}{|p{1cm}|p{8cm}|p{8cm}|}
\hline
&Criterium & Motivation\\
\hline
1)&The Hamiltonian action includes an additional constraint $\phi^\mu$ that reflects the local velocity rotations symmetry 	discussed in the Lagrangian formulation&
This is imposed because the original motivation is to find the meaning of
the local velocity rotations in the Hamiltonian picture\\
\hline
2)&The constraint $\phi^\mu$  generates local symmetry transformation of the action&
A local symmetry is needed in order to justify a later factorization from the PI, similar 
to the known redundant gauge configurations.\\
\hline
3)& The equations of motion are in agreement with the classical equations of the RPP&
Only if the systems are classically equivalent, one is still solving the same problem one was
up for in the first place, namely the quantization of the RPP.\\
\hline
%4) &The new constraint should maintain the existing local and global symmetries of the action&
%%Only if all known existing symmetries are maintained, one can be sure that the physical 
%system has not been changed. This condition seems redundant to the condition 3), 
%but it is listed as extra point because one could think about a replacement of the original constraint
%instead of imposing an additional constraint.\\
%\hline
4)& The PI of the additional constraint can be done and
the result does not modify the expected Klein Gordon propagator&
The factorization of the new symmetry is meant to act as improvement
of the naive PI approach and thus should not introduce modifications where 
this naive approach already works.\\
\hline
\end{tabular}\label{tab2}
\end{table}

%%%%%%%%%%%%%%%
\subsection{Hamiltonian without hidden symmetry: Summary}
The usual Hamiltonian action of the RPP is
\be\label{S0}
S[x,p,n]=\int_{t_1}^{t_f} dt\quad  \left[\dot x \cdot p - n (p^2-m^2)\right],
\ee
where $x^\mu$ is the position variable, $p^\mu$ is the momentum variable,
and $n$ is the Lagrange multiplier imposing the Hamiltonian constraint
\be\label{phi}
H_0=\phi=p^2-m^2.
\ee
 Let's summarize the most important properties of this action.
It is invariant under global transformations $x^\mu \rightarrow x^\mu + \xi^\mu$,
where $\xi^\mu$ is a constant.
It is further invariant under the local transformations generated by (\ref{phi})
\be
\delta x^\mu(\lambda)=\left\{x^\mu(\lambda) ,\int d\lambda' \epsilon(\lambda') \phi(\lambda')\right\}
=2 \epsilon(\lambda) p^\mu(\lambda),
\ee
where the canonical Poisson bracket $\left\{x^{\mu}(\lambda),p_{\nu}(\lambda') \right\} = \delta^{\mu}_{\nu}\,\delta(\lambda-\lambda')$ was used.
The classical equations of motion for (\ref{S0}) are
\bea\label{eom01}
\dot p^\mu &=& 0,\\ \label{eom02} 
\dot x^\mu &=& 2 n p^\mu,\\ \label{eom03}
p^2-m^2&=&0.
\eea
The path integral over (\ref{S0}) that defines the propagator
\be
\langle  x_f^{\mu} - x_1^{\mu} \rangle = \int_{x(t_1)=x_1}^{x(t_f) = x_f} \mathcal{D}[x(t)]\mathcal{D}[p(t)]\mathcal{D}[n(t)]e^{i S[x,p,n]}
\ee
can be obtained from a straight forward calculation. Time discretization is performed as usual over $F$-intervals, such that $\epsilon = (t_f - t_1)/F$
is the size of each time-slice. Hence, for $t_j = t_1 + (j-1)\epsilon$, ($1 \le j \le F$), the discrete coordinates $x(t_j)\rightarrow x_j$, and momenta $p(t_j)\rightarrow p_j$,
with fixed coordinates at the ends $x(t_1) \rightarrow x_1$ and $x(t_f) \rightarrow x_f$, respectively. Similarly, the integration measure over paths becomes, in the
discretised form
\be
\mathcal{D}[x(t)] \mathcal{D} [p(t)] \mathcal{D}[n(t)]\rightarrow \prod_{j=2}^{F-1} d^{d}x_j \prod_{j=1}^{F} d^{d}p_j\, dn_j.
\ee
For example, one can perform the $\int d^{d}x_j$ integrals first, which give delta functions
 in momenta, of the form
$\delta^d(p^\mu_{j+1}-p^\mu_j)$. These delta functions allow to perform all the  momenta integrals $\int d^{d}p_j$,
except for the final one.
The integrals over the Lagrange multipliers of the constraint $\int dn_j$, would give a delta function $\delta (p^2_j-m^2)$ each. In order
to avoid this piling up of delta functions, one fixes all but one $n_j$, thus giving the expected propagator
\bea\label{letzter0}
\langle  x_f^\mu-x_1^\mu \rangle &=&{\mathcal{N}} 
 \int d^dp_F \,
 e^{-i {\left( x_f - x_1\right)}\cdot  p_F} \,\delta(p_F^2-m^2).
\eea

%%%%%%%%%%%%%%%%
\subsection{Ansatz for the constraint}
%In gravity one has some problems: the first problem is that the equation of motion
%for the momentum is not zero $\dot \Pi^{ij}\neq 0$, the second problem is that one can not
%choose an initial condition $g_{ij}=0$, actually it does not make sense to write the constraint
%as $\sim \dot g_{ij} \int_t^{t_f} \Phi^{ij}$.
%Thus, one is forced to write it in the form of $\sim \dot g_{ij} \Phi^{ij}$.
%If such a construction is possible it should also be possible for the point particle, so lets show
%this:

As listed above, one wants a constraint which reflects the velocity rotations in the Lagrangian picture.
Since velocity is a vector, a scalar constraint is insufficient. One needs at least a vector
or tensor for this task. Further,
in the equations of motion (\ref{eom01}-\ref{eom03}) the velocity is associated to
the momentum, thus one might first attempt to formulate a constraint which 
transforms the momenta $p^\mu$.
In order to transform the momenta with a Poisson bracket with $\phi^\mu$, one needs
this constraint to depend on the positions $\phi^\mu=\phi^\mu(x)$.
However, when working out the equations of motion and algebra, 
it becomes clear that such a position dependent
constraint would have to be non-local in the position variables. This can be done, but
we prefer to avoid the problems that come along with non-locality, and hence we search for a constraint
that is local in momentum space but independent of $x^\nu$, namely $\phi^\mu=\phi^\mu(p)$.
Still a change in the momentum can be achieved but in a different way,
as will be seen.

The action with the new constraint is then
\be\label{S1}
S[x,p,n,N]=\int_{t_1}^{t_f} dt\quad  \left[\dot x \cdot p - n (p^2-m^2)- \dot x_\mu \phi^\mu\right].
\ee
Here, we propose
\be\label{consrpp}
\phi^\mu= N^\mu \left(\frac{\delta H_0}{\delta p}\right)^2-\left(\frac{\delta H_0}{\delta p}\right)^\mu \left(N\cdot \left(\frac{\delta H_0}{\delta p}\right)\right) =N^\mu p^2-p^\mu (N\cdot p),
\ee
where $N^\mu$ is the new Lagrange multiplier function that must vanish at the endpoints
and $H_0=\phi$ is the usual Hamiltonian constraint.
This form of the constraint has the useful property that it is by construction orthogonal to the momentum,
namely
\be\label{orto}
p_\mu \phi^\mu=0,
\ee
which will lead to very useful simplifications.
%%%%%%%%%%%%%%%%%%
\subsection{Equations of motion}
One can derive the equations of motion for this system in the variables
$x^\mu$, $p^\mu$, and $N^\mu$. Then, we find that $N^\mu \sim p^\mu$
and thus that the equations of motion are equivalent to the usual ones
\footnote{
It is important to realize that this classical reduction to the usual equations
of motion fails for the non-relativistic theory, because in this theory
the classical relation between $\dot{\vec x}$ and $\vec p$ is fixed and
an arbitary $\vec N$ which is just proportional to $\vec p$ would break this relation.}.
This can be seen more elegantly if one notes that $p^\mu$ is not the canonical momentum $\pi^\mu$
any more since
introducing (\ref{consrpp}) gives
\be
\pi^\mu=p^\mu-\phi^\mu.
\ee
This is the change in the momentum variable we have been seeking.
Thus, in order to derive the equations of motion for the system with $\pi^\mu$,
one would like to rewrite the whole action in terms of $\pi^\mu$ instead of $p^\mu$.
By using (\ref{orto}) one finds that, after the rescaling  $ \tilde N ^\mu=N^\mu \sqrt{n \pi^2}$ ,
the action (\ref{S1}) can be written as
\be\label{S2}
S[x,\pi,n,\tilde N]=\int_{t_1}^{t_f}dt \left[  \dot{x} \cdot \pi - n (\pi^2-m^2)- \tilde N \cdot \tilde \phi\right],
\ee
where
\be\label{consrpp2}
\tilde\phi^\mu =\tilde N^\mu \pi^2-\pi^\mu (\tilde N\cdot \pi ).
\ee
After a variation with respect to $\delta x^\mu$, $\delta \pi^\mu$, $\delta n$, and $\delta \tilde N^\mu$,
the equations of motion are respectively
\bea\label{eom11}
\dot \pi^\mu &=& 0,\\ \label{eom12} 
\dot x^\mu &=& 2 n \pi^\mu+ 2 (\pi^\mu \tilde N^2- \tilde N^\mu (\tilde N \cdot \pi)),\\ \label{eom13}
\pi^2-m^2&=&0,\\ \label{eom14}
\tilde{\phi}^{\mu} &=& 0 = \tilde N^\mu \pi^2 - \pi^\mu (\tilde N \cdot \pi).
\eea
Note that, in order to  compare this with (\ref{eom01}-\ref{eom03}), one should now relabel $\pi^\mu \rightarrow p^\mu$.
One notes then, that the equations (\ref{eom01}) and (\ref{eom03}) are unchanged with respect to
(\ref{eom11}) and (\ref{eom13}).
The equation (\ref{eom12}) acquired an additional term with respect to (\ref{eom02}).
However, this modification vanishes on-shell due to the new equation  (\ref{eom14}),
which forces $\tilde N^\mu$ to be parallel to $\tilde \phi^\mu$.
Thus, the extended action (\ref{S2}) is equivalent to the action (\ref{S0})
at the classical level.
Also all classical symmetries of (\ref{S0}) are present in (\ref{S2}), just as required.

%%%%%%%%%%%%%%%%%%%%
\subsection{The algebra of local transformations}
The term $\tilde \phi= \tilde N_\mu \cdot \tilde \phi^\mu$ generates  local transformations.
The algebra (with $\partial/ \partial\pi^i$ instead of $\partial/ \partial p^i$) is simply
\be
\left\{\tilde \phi(t),\tilde \phi(t') \right\}=0.
\ee
Canonical momenta are unchanged under this constraint
\be
\left\{\pi^\mu(t),\tilde \phi(t') \right\}=0,
\ee
but it
generates a local transformation of the position variables
\be
\delta x^\mu=\left\{x^\mu(t),\int_{t_1}^{t_f} dt'\tilde \phi(t') \right\}=2(\tilde N^2 \pi^\mu- (\tilde N\cdot \pi)N^\mu).
\ee
One realizes that this transformation is orthogonal to the instantaneous momentum direction
\be
\delta x \cdot \pi=0.
\ee
Thus, $\delta x^\mu$  reflects the spirit of the velocity rotations in the Lagrangian formulation~\cite{Koch:2017bvv,Koch:2017nha}. The constraint
further generates a variation of the action
\bea\nonumber
\delta_{\tilde \phi}S&=& \int_{t_1}^{t_f} dt \left[ \delta (\dot x \cdot \pi)- \delta \tilde \phi \right]\\
&=&\int_{t_1}^{t_f} dt \left[ \frac{d}{dt}\left(\tilde  N^2 \pi^2-(\tilde N \cdot \pi)^2\right)
-2 ( \tilde N \cdot (\dot {\tilde N}+\delta \tilde N)) \pi^2+2 (\tilde N \cdot \pi) ((\dot{ \tilde N}+\delta \tilde N) \cdot \pi)\right].
\eea
This variation reduces to a boundary term if the Lagrange multiplier transforms as
\be
\delta \tilde N^\mu=-\dot{ \tilde  N}^\mu.
\ee
Thus, $\tilde \phi$ generates a local symmetry of the action $S$.
%%%%%%%%%%%%%%%%%%
\subsection{Path integral}

For the PI one uses $N^\mu$ without rescaling, which in the discretized version reads
\be\label{actRdis}
S[x^\mu, \pi^\nu, n, N^\alpha]= \sum_{i=1}^{F}\epsilon \left(\left(  \pi_i \cdot\left( x_{i+1}-x_{ i}\right)\right)/\epsilon
-n_i (\pi_i^2-m^2)-n_i \pi_i^2 (N_i\cdot \phi_i)\right).
\ee
The functional integrals are
\bea\label{integrals1}
\int{\mathcal{D}x}&=& \Pi_{i=2}^{F-1} \int d^dx_i\\ \label{integrals2}
\int{\mathcal{D}\pi}&=& \Pi_{i=1}^F \int d^d\pi_i\\ \label{integrals3}
\int{\mathcal{D}n}&=& \Pi_{i=1}^F \int dn_i\\ \label{integrals4}
\int{\mathcal{D}N}&=& \Pi_{i=1}^{F-1} \int d^{d-1}N_{o,i}\cdot \Omega_i.
\eea
Here, we have split the vectors $N^\mu_i$ into a part orthogonal to $\pi^\mu_i$
and a part parallel to $\pi^\mu_i$
\be
N^\mu_i=N_{o,i}^\mu+N^\mu_{i,||}.
\ee
Clearly, in the PI one should only integrate over the orthogonal part $N_{o,i}^\mu$,
since the parallel part does not contribute to the action (\ref{S2}).
The properly chosen measure for this integration  $ \int d^{d-1}N_{o,i}$ is
 $\Omega_i$. Note that this adjustment of the measure is not unusual, since it is known that canonical transformations, similar to those generated by $N_{o,i}^\mu$, can result in a change of the
 path integral measure~\cite{Swanson:1994jn,Blasone:2017spx,Blasone:2017wqa}.

As mentioned after (\ref{S1}) we further impose that the final $N_{o,F}^\mu=0$.
Let us perform the integrals (\ref{integrals1}-\ref{integrals4}) blockwise for each time step.
The first block contains $\int d^dx_2 \int d^d\pi_1\int dn_1\int d^{d-1}N_1 \Omega_1$
which will now be integrated in the same order
\bea\label{erster}
\langle  x_f^\mu-x_1^\mu \rangle&=& \left(\int d^dx_2 \int d^d\pi_1 \int dn_1 \int d^{d-1} N_{o,1}   \right)
e^{i\left[ \left((x_2-x_1) \cdot \left( \pi_2 -  \pi_{1} \right)\right)- \epsilon n_1 ( \pi_1^2(1+N_1^2)-m^2)\right]} \dots.
\eea
Now, integrating over $\int d^dx_2$ gives a $\delta$-function in $\pi^\mu$, which can be used to integrate
$ \int d^d\pi_1$.
Then integrating in $\int dn_1$ gives another  $\delta$-function
\be
\delta(\pi_2^2 (1+ N_{o,1}^2)-m^2),
\ee
which can be used to evaluate the ``radial'' part of the 
$\int d^{d-1} N_{o,1}$ integral (the angular part is just normalization).
For a proper choice of the measure $\Omega_1$, all integrals in this set have canceled each other
and the contribution is just a multiplicative one.
This procedure continues until the final integrals, which have no
$\int d^dx_{F}$ and no $\int d^{d-1} N_{o,F}$, and which thus read
\bea\label{letzter}\nonumber
\langle  x_f^\mu-x_1^\mu \rangle&=&{\mathcal{N}}' \cdot \int d^d\pi_F \int dn_F \,
 e^{-i \left( (x_f -x_1)\cdot  p_F - \epsilon n_F (\pi_F^2-m^2) \right)} \\
 &=&{\mathcal{N}} 
 \int d^d\pi_F \,
 e^{-i \left( (x_f-x_1) \cdot  \pi_F\right)} \,\delta(\pi_F^2-m^2),
\eea
where ${\mathcal{N}}'$ and ${\mathcal{N}} $ are just normalization constants.
This is the desired usual result (\ref{letzter0}).

It is very interesting to note, that in this PI construction the integrals
of the two constraints canceled each other and thus,
we did not
even have to fix some gauges explicitly as in the usual case.
This non-trivial impact on the PI formulation gives further evidence, that the new constraint,
even though ``hidden'' at the classical level, 
is not ``trivial'' at the level of the quantum mechanical path integral formulation.

%%%%%%%%%%%%%%%%%%%%%%%%
\section{summary and outlook}

We have presented a new constraint (\ref{consrpp}) for the RPP and shown that
it reflects all the nice features demanded in table \ref{tab2}.
In particular, it generates a non-trivial local symmetry of the action 
without altering the classical equations of motion of the system.
This new  PI formulation worked out in a straight forward way.
Based on our results, we can conclude that the local symmetry of the Lagrangian version
of the RPP action (\ref{actL}), which was discussed in~\cite{Koch:2017bvv,Koch:2017nha} can
also be implement in the Hamiltonian version (\ref{S2}). 

This seems to be a very isolated result, only valid for a very particular system
with general covariance. 
However, as recently shown by Hojman,
there is a very large class of point particle systems which can be cast
in the form of the free RPP which is living on a higher dimensional Rindler background~\cite{Hojman:2018jfe,Hojman:2018mci}.
Thus, we believe that our results might be applicable to a much larger class of problems.
Further, it would be interesting to explore certain similarities of the presented constraint with
 constraints imposed in delta-theories such as delta gravity~\cite{Alfaro:2013ega}.
Finally, our result encourages further investigation on more complicated covariant systems
such as quantum cosmology~\cite{Vilenkin:1994rn} or ultimately quantum gravity in the canonical formulation~\cite{Arnowitt:1960es} analogous
to (\ref{S0}/\ref{S2}).

\section*{Acknowlegements}

B.~K. was supported by  Fondecyt 1161150 and Fondecyt 1181694. 
E.~M. was supported by  Fondecyt Regular 1190361.

\end{document}